\documentclass[a4paper]{article}

\usepackage{INTERSPEECH2021}
\usepackage{hyperref}
\usepackage{multirow}

\title{CVC: Contrastive Learning for Non-parallel Voice Conversion}
\name{Tingle Li$^{1,3,*}$, Yichen Liu$^ {1,*}$, Chenxu Hu$^{1,2,*}$\thanks{\hspace{-1mm}* Equal Contribution}, Hang Zhao$^{1,3,\dagger}$\thanks{\hspace{-1mm}$\dagger$ Corresponding Author}}
\address{
$^1$Institute for Interdisciplinary Information Sciences, Tsinghua University, China\\
$^2$College of Computer Science and Technology, Zhejiang University, China\\
$^3$Shanghai Qi Zhi Institute, China
}
\email{hangzhao@mail.tsinghua.edu.cn}

\begin{document}

\maketitle
\begin{abstract}
Cycle consistent generative adversarial network (CycleGAN) and variational autoencoder (VAE) based models have gained popularity in non-parallel voice conversion recently. However, they often suffer from difficult training process and unsatisfactory results. In this paper, we propose CVC, a contrastive learning-based adversarial approach for voice conversion. Compared to previous CycleGAN-based methods, CVC only requires an efficient one-way GAN training by taking the advantage of contrastive learning. When it comes to non-parallel one-to-one voice conversion, CVC is on par or better than CycleGAN and VAE while effectively reducing training time. CVC further demonstrates superior performance in many-to-one voice conversion, enabling the conversion from unseen speakers.

\end{abstract}
\noindent\textbf{Index Terms}: Non-parallel Voice Conversion, Contrastive Learning, Noise Contrastive Estimation

\section{Introduction}
\label{sec:intro}
In voice conversion (VC), given a speech signal from a source speaker, we aim to convert it to the voice of a target speaker while preserving the speech content. Undoubtedly, developing VC method requires speech data from both speakers. Early methods, such as \cite{early_work_1,early_work_2,early_work_3,early_work_7,early_work_8}, were typically developed using fully or partially parallel speech data, \textit{i.e.}, the source and target speech pairs share the same linguistic content and are temporally aligned. However, the availability of parallel speech pairs is very limited due to the humongous cost in data collection, therefore more recent studies are focused on non-parallel VC. 

Non-parallel VC can be formulated as a regression problem of approximating a mapping function from source speaker domain to target speaker domain, \textit{i.e.}, a domain-to-domain transformation. Cycle consistent adversarial network (CycleGAN) \cite{CycleGAN2017} was first introduced to tackle unpaired image translation problem which is comparable to non-parallel VC since both tasks require domain-to-domain transformation. It follows that several state-of-the-art works in non-parallel VC, such as CycleGAN-VC~\cite{CycleGAN-VC,CycleGAN-VC2, kaneko2020CycleGAN-VC3} and StarGAN-VC~\cite{StarGAN-VC2,StarGAN-VC}, are based on the cycle-consistency mechanism, where a consistency loss has been applied to construct an invertible mapping such that the generated speech can sufficiently preserve the speech content from the source speaker. 

Although cycle consistency is an effective mechanism for controlling the generation process, it is often too restrictive in real-life scenarios. Specifically, it introduces a strict ``pixel-level" constraint that forces a model to be a bijective mapping, such that it cannot ignore irrelevant features during the conversion process. In VC, this disadvantage can cause the converted voice still retaining much of the voice style of the source speaker. As a support of our argument, \cite{zhao2020unpaired} has also pointed out the same issue when using cycle consistency loss in unpaired image-to-image translation problems. Moreover, it is known that GANs are notoriously difficult to train. The situation could be worse when training CycleGAN models which consist of two GANs in order to accomplish the inverse mapping. Thus, developing such models may involve great training cost and extra tuning.

Besides the adversarial approaches mentioned above, there are also works based on auto-encoder (AE) and variational auto-encoder (VAE). AutoVC~\cite{autoVC} used AE to construct a suitable bottleneck that allows the network to separately encode content information into latent representations. The solution presented in~\cite{hsu2016voice} applies a regular VAE, along with latent sampling and the Gaussian reparameterization trick~\cite{VAE}, which is one of the early attempts on non-parallel VC. In more recently emerged works, VAE has been combined with contrastive predictive coding (CPC) \cite{Representation_learning_CPC} or cycle consistency loss (CycleVAE)~\cite{tobing2019non} to tackle voice conversion.
In \cite{VAE_CPC}, CPC is used as an additional regularization objective for enhancing the content encoder.
VQVAE-CPC \cite{VQVAE_CPC} used vector-quantized VAE (VQ-VAE) \cite{VQ-VAE_paper} where the quantized latent representations could discard undesired attributes from the source speech more effectively. 

Inspired by the recent work \cite{NCE_image} for unpaired image translation, we propose a concise model for voice conversion, \textit{i.e.}, CVC, which contains two explicit training objectives, namely the adversarial and contrastive losses. Compared to CycleGAN-VC3 \cite{kaneko2020CycleGAN-VC3}, CVC only requires a regular one-way GAN structure, which greatly reduces the training difficulty of GAN. Besides, in contrast to the previous CPC-VAE based approaches, we aim to establish correspondence between source and target content, by applying contrastive losses at different levels (scales) of the spectral features. According to the experimental results, both subjective and objective evaluation of naturalness and similarity show that for each testing pair of voices, CVC achieves on par or better performance compared to the two baseline models: CycleGAN-VC3 \cite{kaneko2020CycleGAN-VC3} (for notational convenience, we denote by CycleGAN-VC later) and VAE-VC \cite{hsu2016voice}.

The remainder of this paper is organized as follows. In Section \ref{sec:method}, we explain the details of our proposed method. Experimental setup and results are reported in Section \ref{sec:experiments}, while a concise summary and description of future work are presented in Section \ref{sec:conclusion}.

\section{Contrastive Voice Conversion}
\label{sec:method}
Under the unsupervised voice conversion learning setting, the general goal is to learn a spectral mapping from source domain $\mathcal{X}$ to target domain $\mathcal{Y}$ without using parallel speech pairs. Here, each domain can be defined as either a single speaker or multiple speakers treated as an entity. Our proposed CVC is capable of conducting both one-to-one and many-to-one voice conversion. This can be achieved through two distinct training objectives, which will be explained in this section.

\subsection{Adversarial Training}
In contrast to the previous CycleGAN-based approaches, CVC only requires a single GAN, which largely simplifies the training process. The generator network $G$ can be divided into two components, an encoder $G_{\text{enc}}$ followed by a decoder $G_{\text{dec}}$. For a given dataset of non-parallel speech instances $X=\{\bm{x}\in\mathcal{X}\}$, $Y=\{\bm{y}\in\mathcal{Y}\}$, $G_{\text{enc}}$ and $G_{\text{dec}}$ are applied sequentially to generate the output spectrogram $\hat{\bm{y}}=G_{\text{dec}}(G_{\text{enc}}(\bm{x}))$. An adversarial loss \cite{adversarial_loss} is then applied to encourage $\hat{\bm{y}}$ to approach the spectral features of the target domain $\mathcal{Y}$:
\begin{equation}\label{eq: adversarial_loss}
\begin{split}
    \mathcal{L}_{\text{GAN}}(G_{X\to Y},D_Y)= \text{ }&\mathbb{E}_{\bm{y}\sim Y}\log{D(\bm{y})} + \\
    &\mathbb{E}_{\bm{x}\sim X}\log{(1-D(G(\bm{x})))}
\end{split}
\end{equation}
where $D$ is the discriminator network. 

\subsection{Establishing Mutual Correspondence via Contrastive Learning}
\begin{figure}[t]
\centering
\includegraphics[width=\linewidth]{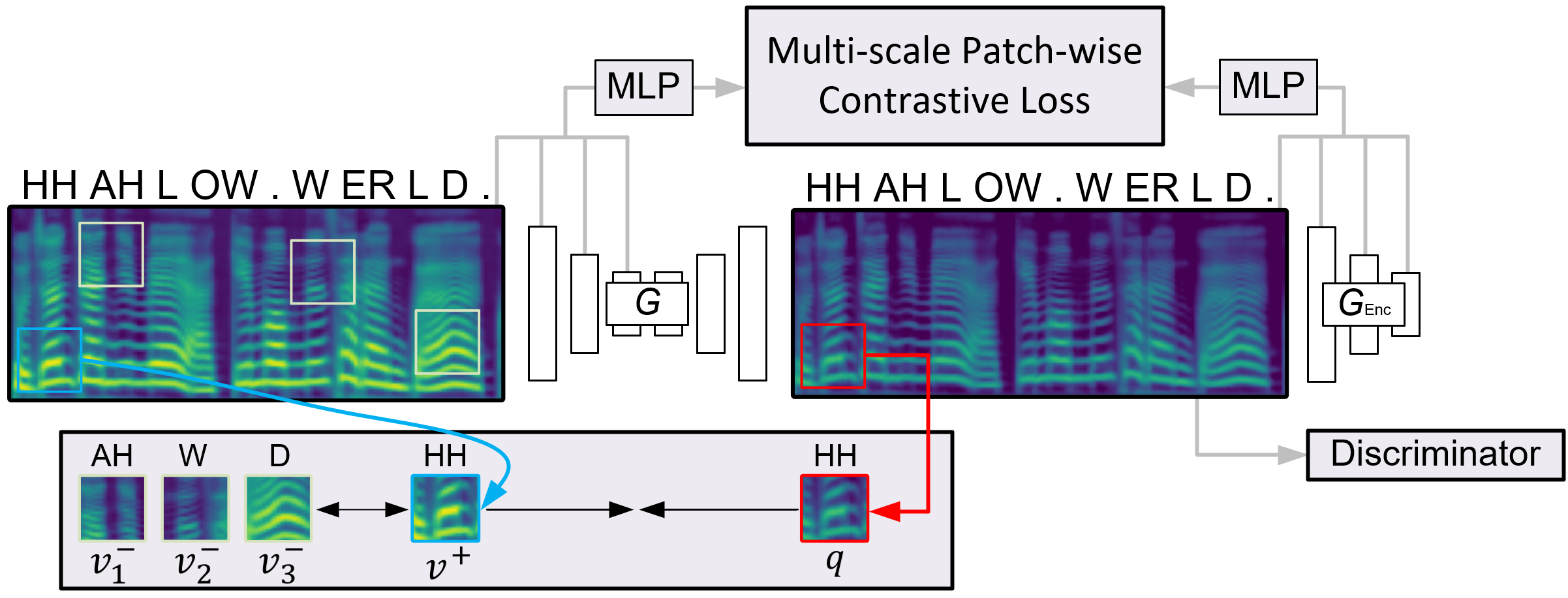}
\caption{Patch-wise contrastive learning for one-way voice conversion. The phoneme type of ``hello world" is taken as an example since we aim to preserve the content information among other features. A generated spectral patch (red) representing ``HH" should match its corresponding input patch (blue) ``HH", in comparison to other random patches ``AH", ``W" or ``D". We use a multi-scale patch-wise contrastive loss \cite{NCE_image}, which establishes mutual correspondence between source and generated spectrograms. This enables one-way voice conversion using non-parallel corpora. Note that the MLP component has been ignored during inference.}
\label{fig: show_NCE}
\end{figure}

In VC, a successfully converted speech should be equipped with the target speaker style while fully preserving the content of the source speech. However, both information, \textit{i.e.}, content and speaker information, are inherently entangled within the spectral feature, and adversarial training can only guarantee style transfer in the results. One trivial solution is that we get a same target speaker audio for any input. Therefore, as shown in Figure~\ref{fig: show_NCE} we introduce the second training objective, based on noise contrastive estimation (NCE) \cite{NCE}, which aims to preserve content information by establishing correspondence between the source and generated spectrograms, $\bm{x}$ and $\hat{\bm{y}}$ respectively.
Note that this training objective is only employed to the encoder network $G_\text{enc}$, which is a multi-layer convolutional network that transforms the source spectrogram into feature stacks at different levels. 
In this way, we encourage $G_\text{enc}$ to throw away timbre information during the transformation, and preserve all content information; then the job of the decoder network $G_\text{dec}$ is to add to the speech the style of the target speaker.

Given a ``query" vector $\bm{q}$, the fundamental objective in contrastive learning is to optimize the probability of selecting the corresponding ``positive" sample $\bm{v}^+$ among $N$ ``negative" samples $\bm{v}^-$. The query, positive and $N$ negatives are transformed to $M$-dimensional vectors, \textit{i.e.}, $\bm{q}$, $\bm{v}^+\in\mathbb{R}^M$ and $\bm{v}^-\in\mathbb{R}^{N\times M}$. This problem setting can be expressed as a multi-classification task with $N+1$ classes:
\small
\begin{equation}\label{eq: NCE_expression}
    \ell(\bm{q},\bm{v}^+,\bm{v}^-)=-\log \left(\frac{\exp(\frac{\bm{q}\cdot\bm{v}^+}{\tau})}{\exp(\frac{\bm{q}\cdot\bm{v}^+}{\tau})+\Sigma_{n=1}^N\exp(\frac{\bm{q}\cdot\bm{v}^-_n}{\tau})} \right) 
\end{equation}
\normalsize
where $\bm{v}^-_n$ denotes the n-th negative sample and $\tau$ is a temperature parameter, as suggested in SimCLR \cite{SimCLR}, which scales the similarity distance between $\bm{q}$ and other samples. The crossentropy term in \eqref{eq: NCE_expression} represents the probability of matching $\bm{q}$ with the corresponding positive sample $\bm{v}^+$. Thus, iteratively minimizing the negative log-crossentropy is equivalent to establishing mutual correspondence between the query space and the sample space. 

In our VC task, we draw the $N+1$ positive/negative samples internally from the source spectrogram $\bm{x}\in X$, and the query $\bm{q}$ is selected from the generated spectrogram $\hat{\bm{y}}$. From Figure \ref{fig: show_NCE}, it can be seen that the selected samples are treated as ``patches" that capture local information among the spectrograms. This setup is motivated by the logical assumption that the global correspondence between $\bm{x}$ and $\hat{\bm{y}}$ is determined by the local, \textit{i.e.}, patch-wise, correspondences. In other words, patch-wise optimization on multi-scale features would eventually lead to the convergence from local to global correspondence. 

Since the encoder $G_\text{enc}$ is a multi-layer network that maps $\bm{x}$ into feature stacks after each layer, we choose $L$ layers and pass their feature stacks through a small MLP network $P$. The output of $P$ is $P(G_\text{enc}^l(\bm{x}))=\{\bm{v}_l^1,...,\bm{v}_l^N,\bm{v}_l^{N+1}\}$, where $l\in \{1,2,...,L\}$ denotes the index of the chosen encoder layers and $G_\text{enc}^l(\bm{x})$ is the output feature stack of the $l$-th layer. Similarly, we can obtain the query set by encoding the generated spectrogram $\hat{\bm{y}}$ into $\{\bm{q}_l^1,...,\bm{q}_l^N,\bm{q}_l^{N+1}\}=P(G_\text{enc}^l(\bm{\hat{y}}))$. Now we let $\bm{v}_l^n\in\mathbb{R}^M$ and $\bm{v}_l^{(N+1)\text{\textbackslash}n}\in \mathbb{R}^{N\times M}$ denote the corresponding positive sample and the $N$ negative samples, respectively, where $n$ is the sample index and $M$ is the channel size of $P$. By referring to Eq. \eqref{eq: NCE_expression}, our second training objective can be expressed as:
\begin{equation}\label{eq: NCE_objective}
    \mathcal{L}_\text{NCE}(G_\text{enc},P,X)=\mathbb{E}_{\bm{x}\sim X} \sum_{l=1}^{L} \sum_{n=1}^{N+1} \ell(\bm{q}_l^n,\bm{v}_l^n,\bm{v}_l^{(N+1)\text{\textbackslash}n})
\end{equation}
which is the average NCE loss from all $L$ encoder layers.

\subsection{Overall objective}
In addition to the two objectives discussed above, we have also employed an identity loss $\mathcal{L}_\text{identity}=\mathcal{L}_\text{NCE}(G_\text{enc},P,Y)$ which utilizes the NCE expression in Eq. \eqref{eq: NCE_objective}. By taking the NCE loss on the identity generation process, \textit{i.e.}, generating $\hat{\bm{y}}$ from $\bm{y}$, we are likely to prevent the generator from making unexpected changes. Now we can define our final training objective as:
\begin{equation}\label{eq: final_objective_loss}
\begin{split}
    \mathcal{L}_\text{CVC}= \text{ }&\mathcal{L}_{\text{GAN}}(G_{X\to Y},D_Y)+\lambda\mathcal{L}_\text{NCE}(G_\text{enc},P,X)+ \\ &\mu\mathcal{L}_\text{NCE}(G_\text{enc},P,Y)
\end{split}
\end{equation}
where $\lambda$ and $\mu$ are two parameters for adjusting the strengths of the NCE and the identity loss. 

\begin{table*}[tb]
\caption{Comparison objective voice similarity (ranging from 0 to 1) of our proposed CVC with other baseline models on the VCTK dataset, where the best results are shown in bold. For the one-to-one VC, this metric is evaluated using 10 speaker pairs for each gender category, and the many-to-one VC is evaluated with 100 pairs.}
\centering
\begin{tabular}{c|c c c|c c c| c c}
\toprule
\multicolumn{1}{c|}{\multirow{3}*{\textbf{Gender}}}& 
\multicolumn{8}{c}{\textbf{Voice Similarity}}\\
\cline{2-9}
&
\multicolumn{3}{c|}{One to One} & 
\multicolumn{3}{c|}{Many to One} &
\multicolumn{2}{c}{Many (unseen) to One} \\
  &\textbf{CVC} & CycleGAN~\cite{kaneko2020CycleGAN-VC3} & VAE~\cite{hsu2016voice} & \textbf{CVC} & CycleGAN~\cite{kaneko2020CycleGAN-VC3} & VAE~\cite{hsu2016voice} & \textbf{CVC} & CycleGAN~\cite{kaneko2020CycleGAN-VC3}\\
\midrule
Male-Male               & \textbf{0.964} & 0.961    & 0.816 & \textbf{0.947}          & 0.941     & 0.778 & \textbf{0.992} & 0.988      \\
Male-Female             & \textbf{0.979} & 0.964    & 0.805 & \textbf{0.942}          & 0.926     & 0.803 & \textbf{0.952} & 0.935      \\
Female-Female           & \textbf{0.965}          & 0.937    & 0.828 & \textbf{0.966} & 0.945     & 0.831 & \textbf{0.963} & 0.958      \\
Female-Male             & \textbf{0.983} & 0.979    & 0.874 & \textbf{0.981}          & 0.979     & 0.845 & \textbf{0.988} & 0.984      \\
\bottomrule
\end{tabular}
\label{tb: voice_similarity}
\end{table*}

\section{Experiment}
\label{sec:experiments}

In this section, we will introduce the training schemes of our proposed model which was evaluated on one-to-one and many-to-one voice conversion experiments. Besides, some auditory demos are available online\footnote{\url{https://tinglok.netlify.com/files/cvc/}}.

\subsection{Experimental Setup}
\textbf{Dataset.} VCTK corpus \cite{veaux2016superseded}, which contains 44 hours of utterances from 109 speakers, was used to evaluate our proposed model. It is a suitable dataset for fully non-parallel VC setting because each speaker reads different sentences, except the rainbow passage and the elicitation paragraph (we already ruled out these utterances before training). For the one-to-one VC experiment, utterances from two different speakers, \textit{i.e.}, the source and target speaker, were used for training. And 50 utterances from the source speaker, which were disjoint to the training utterances, were selected for model evaluation. Moreover, the many-to-one VC experiment involved multiple source speakers which were treated as a single source domain. Similar to the one-to-one VC experiment, 50 novel utterances from each source speaker were selected for evaluation. Please note that the utterances from 9 chosen speakers were excluded in this experiment, in order to use them to also evaluate the model performance on unseen source speakers after training. 

\noindent \textbf{Conversion process.} The pretrained Parallel WaveGAN (PWG) vocoder\footnote{\url{https://github.com/kan-bayashi/ParallelWaveGAN}} \cite{yamamoto2020parallel} was used for synthesizing the waveform signals from the converted log-Mel-spectrogram.

\noindent \textbf{Network architecture.} The encoder and decoder of CVC's generator are fully convolutional, with 9 layers of ResNet-based CNN bottlenecks \cite{he2016deep} in between. All the kernel size of CNN is set to $3 \times 3$, and the stride size depends on whether downsampling is required. For the discriminator, we applied the PatchGAN architecture \cite{isola2017image} as used in CycleGAN-VC. Please note that all the padding types were modified to replication padding rather than zero padding or reflection padding, in order to circumvent the trivial solution and collapse. 

\noindent \textbf{Training setting.} We used a fix duration of 2s for training efficiency, and those less than 2s were taken away because zero padding would introduce tremendous redundancy and cause contrastive loss difficult to converge (the same reason as above). Before training, all of the utterances were first converted to 24 kHz and 32-bit precision in floating-point PCM format. Then, we extracted 80-dimensional log-Mel-spectrogram features from them, with 25 ms frame length and 10 ms frame-shift, and used a frame-level energy-based voice activity detector (VAD) to filter out non-speech frames. Finally, our proposed CVC was trained with a batch size of 1 and an initial learning rate of 2e-4 for 1000 epochs, using the Adam optimizer. The weights in Eq. \eqref{eq: final_objective_loss} were set to $\lambda = 1, \mu = 1$. Further training schemes and hyperparameter settings are released along with the code\footnote{\url{https://github.com/Tinglok/CVC}}.

\subsection{Evaluation Metrics}
We performed two subjective tests on Amazon Mechanical Turk (MTurk)\footnote{\url{https://www.mturk.com/}}. The first test used the mean opinion score (MOS) \cite{loizou2011speech} metric to evaluate the converted utterances from the one-to-one VC experiment. For each utterance, the listeners were asked to rate a score of 1-5 (higher score indicates better quality) on the voice naturalness and a score of 1-4 on the voice similarity, given the target voice. In contrast to the first test, the second test used the comparison mean opinion score (CMOS) metric. This test aimed to evaluated the converted utterances, from the 9 unseen source speakers, in the many-to-one VC experiment. The listeners were asked to distinguish between the utterances generated by CVC and the baseline model, in terms of better naturalness and voice similarity.


Furthermore, the voice encoder system~\cite{wan2018generalized}, which is an objective test, was applied to evaluate the voice similarity between each converted utterance and its corresponding target utterance. Specifically, as done in previous work \cite{jia2018transfer}, we used Resemblyzer\footnote{\url{https://github.com/resemble-ai/Resemblyzer}}, an open-source text-independent speaker verification (SV) system, which is applicable to compute the voice similarity in non-parallel VC. This SV system took a converted utterance as input and generated a fix-dimensional embedding, then the cosine similarity is calculated between the embeddings of the target utterance and the converted utterance. Finally, it yields scores that range from 0 to 1, where 0 corresponds to different speakers with high confidence, and 1 corresponds to the same speaker with high confidence.

\subsection{One-to-one VC Experiment}
\begin{figure}[t]
\begin{minipage}[b]{.48\linewidth}
  \centering
  \centerline{\includegraphics[width=\linewidth]{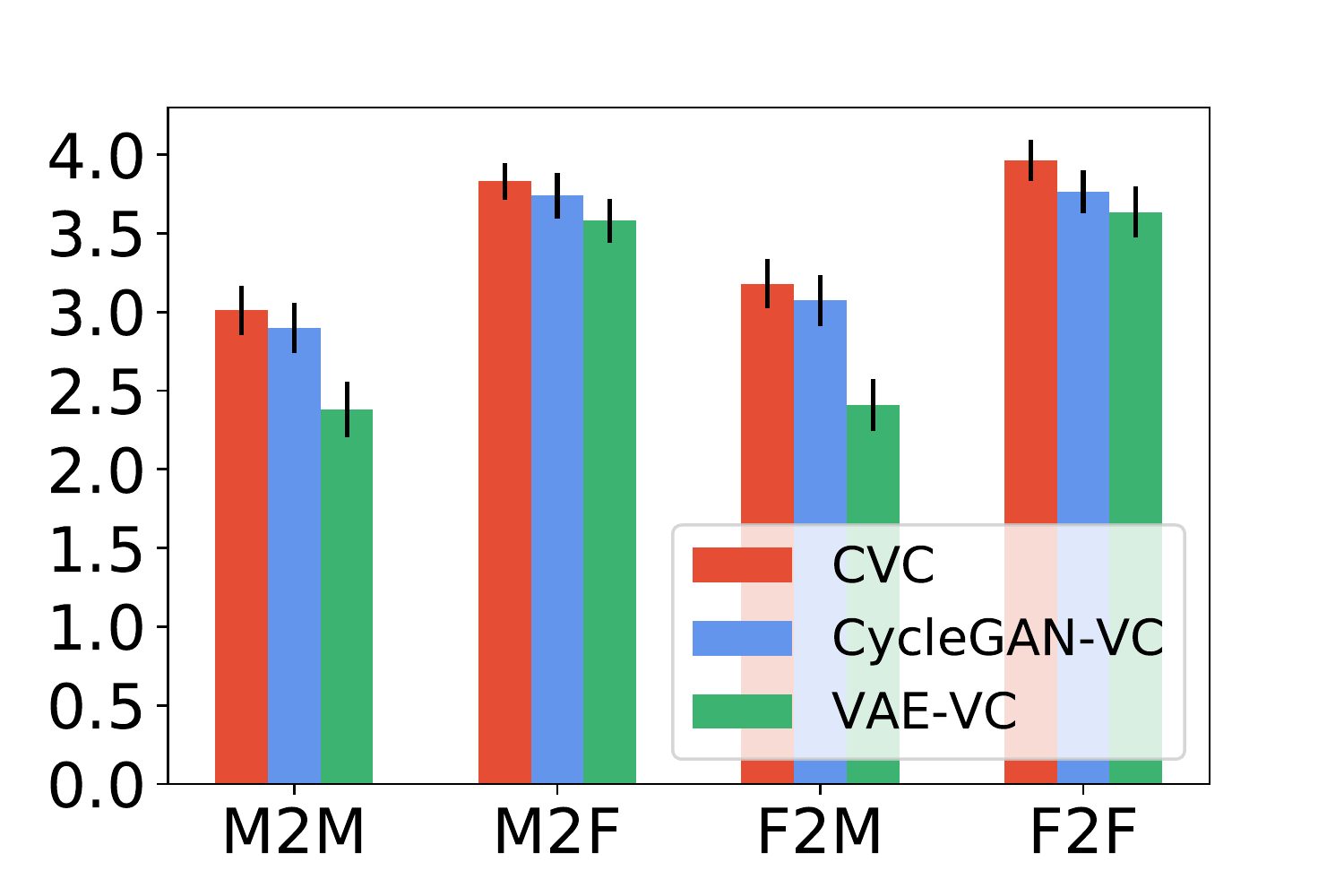}}
\end{minipage}
\hfill
\begin{minipage}[b]{0.48\linewidth}
  \centering
  \centerline{\includegraphics[width=\linewidth]{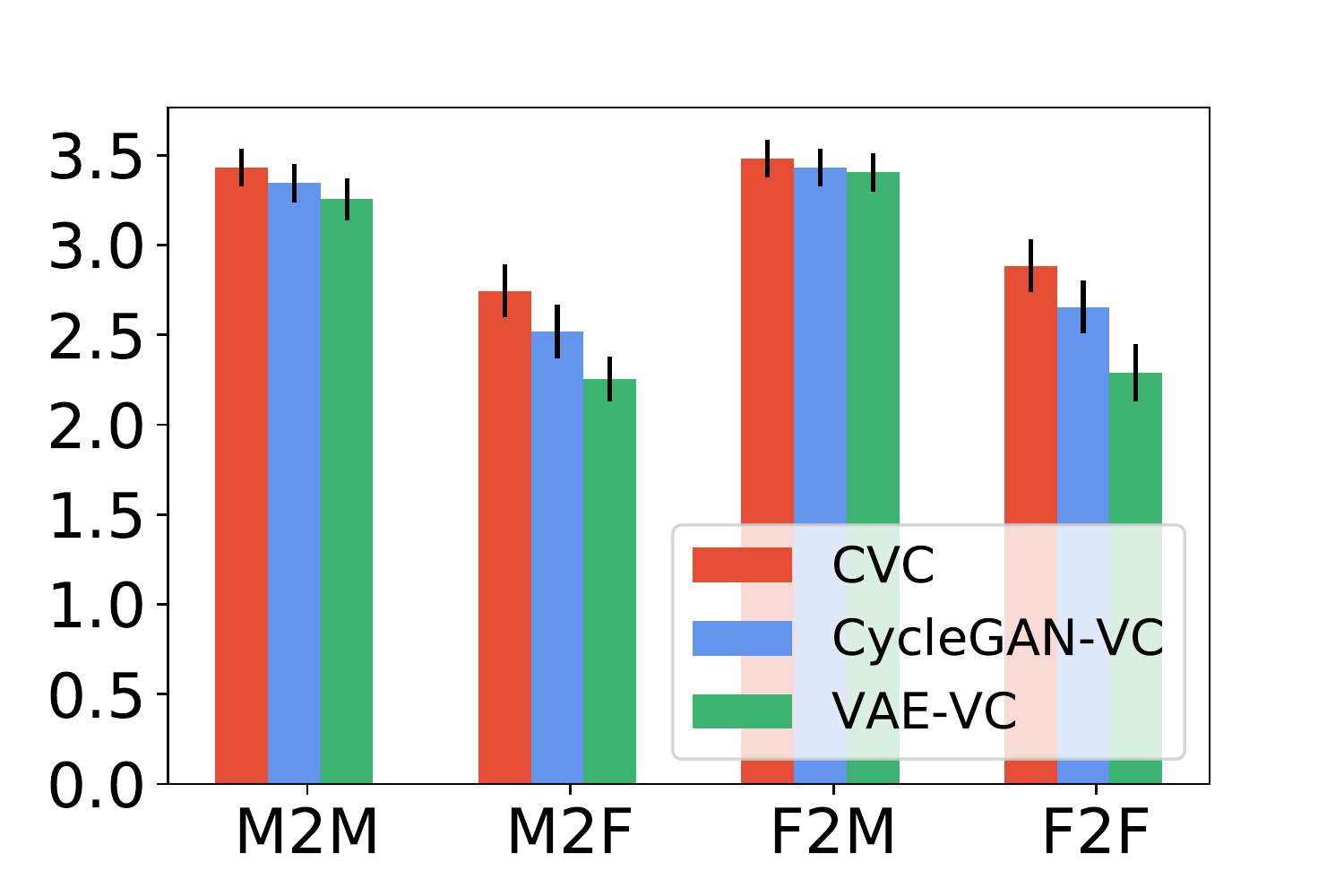}}
\end{minipage}
\caption{MOS results on voice naturalness (left) and similarity (right) for one-to-one VC, where the error bars denote 95\% confidence interval.}
\label{fig:show_mos}
\end{figure}

Here, two state-of-the-art baselines, \textit{i.e.}, CycleGAN-VC \cite{kaneko2020CycleGAN-VC3} and VAE-VC \cite{hsu2016voice}, were compared to our proposed CVC in one-to-one VC task. Please note that in the original work of CycleGAN-VC, MelGAN vocoder \cite{kumar2019melgan} was applied to reconstruct the waveform signals. However, in order to make fair comparison, we changed this setting to PWG vocoder \cite{yamamoto2020parallel} as used in CVC. For VAE-VC, we used the official implementation\footnote{\url{https://github.com/JeremyCCHsu/vae4vc}} and changed the training data to VCTK \cite{veaux2016superseded}. It can be seen from Table~\ref{tb: voice_similarity} that CVC has higher objective voice similarity scores on both inter-gender and intra-gender conversions, which implies CVC generates better speaker timbre. Furthermore, when the target speaker is a female, the score discrepancies between CVC and the baseline models are relatively large, showing that CVC is more suitable for high frequency modeling. Besides, as shown in Figure~\ref{fig:show_mos}, the MOS scores of CVC are also better than the other baseline models, thus indicating that has better content and timbre encoding abilities than the others. Results in both experiments demonstrate that our proposed model yields better one-to-one voice conversion performance.

Besides performance, we also compared the training time between CVC and CycleGAN-VC under the same environmental condition, where a cluster that contains 64 AMD Ryzen Threadripper 2990WX 32-Core Processor CPU and 1 GeForce RTX 2080 Ti GPU was used for training. As the result, the average time elapsed of CVC (518 min) was less than that of CycleGAN-VC (574 min), suggesting that CVC only needs 90.2\% of CycleGAN-VC's training time. This follows our expectation since CVC only uses a regular GAN structure, as opposed to a dual one used in CycleGAN-VC. 


\subsection{Many-to-one VC Experiment}
This experiment consists of two explicit parts. The first part evaluates the performance of CVC on the seen source speakers while the second part evaluates on the 9 unseen source speakers. Please note that only CycleGAN-VC was used as the baseline model in the second part because VAE-VC is not capable of converting utterances from unseen source speakers. The objective similarity scores of the first and second part are shown in the last two columns of Table~\ref{tb: voice_similarity}. Interestingly, when the target speaker is male, the voice similarity score is better than when it is female, suggesting that it may be easier to convert to a low frequency voice than a high frequency voice. Besides, CVC surpasses CycleGAN-VC and VAE-VC in both inter-gender and intra-gender conversions, which manifests that CVC has better timbre conversion capability than the baseline models.

\begin{figure}[t]
\centering
\includegraphics[width=\linewidth]{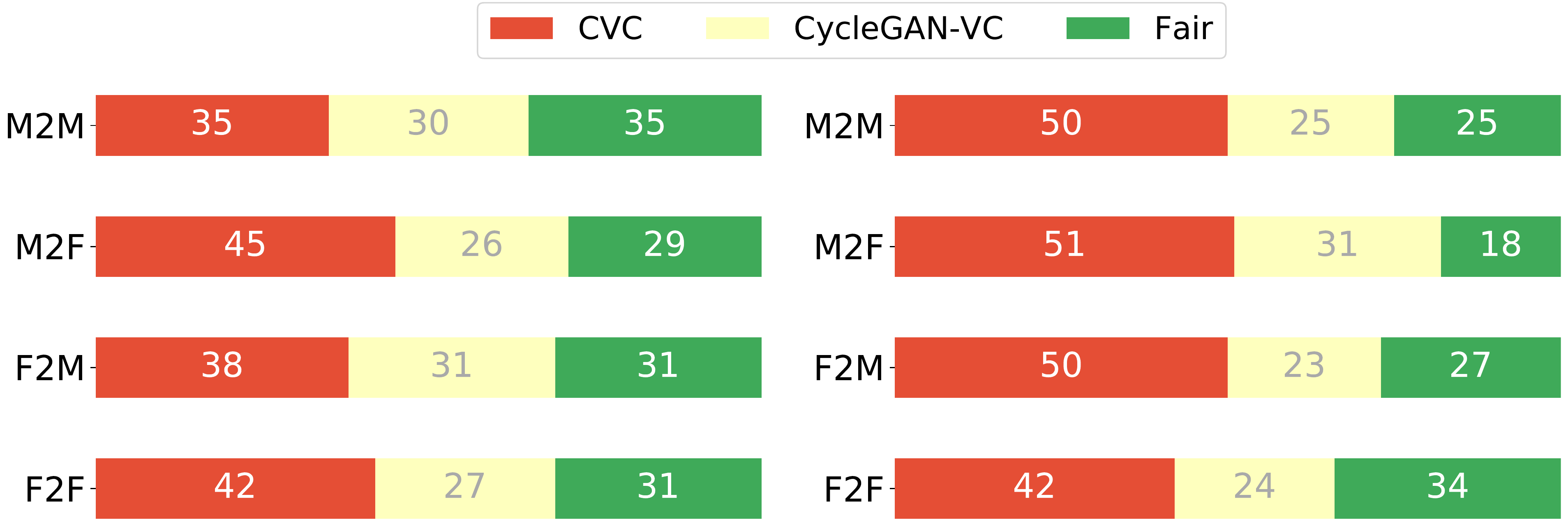}
\caption{CMOS scores (\%) on voice naturalness (left) and similarity (right) for many-to-one VC, where the source speakers are unseen speakers.}
\label{fig: show_cmos}
\end{figure}

Furthermore, according to Figure~\ref{fig: show_cmos}, the naturalness and similarity CMOS of CVC also achieved competitive performance compared to CycleGAN-VC in both inter-gender and intra-gender conversions. Hence, it can be stated that CVC is more effective than the state-of-the-art baseline models in both the objective and subjective metrics.


\subsection{Ablation Experiment}
First, it is trivial to do an ablation experiment that sets both $\lambda$ and $\mu$ in Eq. \eqref{eq: final_objective_loss} to zero, if so CVC only contains a regular GAN structure and will simply collapse. The emergence of CycleGAN could be a support to this statement. Hence, the only ablation we can do is to set $\mu$ to zero to omit the identity mapping loss. We begin with the comparison \textit{w.r.t.} the objective voice similarity. The empirical results are reported in Table~\ref{tb: ablation}, where $\Delta$Imp. denotes the relative improvements of CVC over the one without identity mapping loss. We find that the objective similarity scores of CVC consistently outperforms the one without identity mapping loss, which means this loss is likely to enhance timbre conversion ability. More intriguingly, it seems that the gap of intra-gender conversion is larger than that of inter-gender conversion, indicating this loss may be able to promote timbre conversion, particularly for intra-gender conversion. From our perspective, this loss can regularize the generator to approach an identity mapping when real samples of the target speaker are provided. In other words, if something already sounds like the target speaker, this loss is supposed to discourage the generator to convert the utterance in case it is already in the correct domain. As the result, we confirmed that applying identity mapping loss is relatively important.

\begin{table}[tb]
\caption{Ablation experiment of the identity mapping loss concerning the objective voice similarity. The greatest percentage improvements are shown in bold.}
\centering
\renewcommand{\arraystretch}{1.1}
\begin{tabular}{c c c c c}
\toprule
\multicolumn{1}{c}{\multirow{2}*{\textbf{Gender}}}& 
\multicolumn{2}{c}{\textbf{Voice Similarity}}&
\multicolumn{1}{c}{\multirow{2}*{\textbf{$\Delta$Imp.}}}\\
\cline{2-3}
  &CVC & CVC (w/o idt) \\
\midrule
Male-Male               & 0.964 & 0.952 & \textbf{1.26\%}    \\
Male-Female             & 0.979 & 0.973 & 0.62\%    \\
Female-Female           & 0.965 & 0.954 & \textbf{1.15\%}    \\
Female-Male             & 0.983 & 0.979 & 0.41\%    \\
\bottomrule
\end{tabular}
\label{tb: ablation}
\end{table}

\section{Conclusion}
\label{sec:conclusion}
In this paper, we proposed a concise method, \textit{i.e.}, CVC, for preserving content information in non-parallel VC problems, where contrastive learning is highlighted as an effective mechanism for establishing correspondence between source input and generated output. Experimental results show that CVC yields better subjective and objective scores than the baseline models in both one-to-one and many-to-one VC tasks. For future work, CVC can be simply generalized towards any-to-any VC by adding speaker embeddings to the network, which could be a promising direction. Moreover, we can explore the potential of CVC on one-shot VC task, since the contrastive-based models have shed some light on this area.


\newpage
\bibliographystyle{IEEEtran}
\bibliography{mybib}

\begin{thebibliography}{10}
\providecommand{\url}[1]{#1}
\csname url@samestyle\endcsname
\providecommand{\newblock}{\relax}
\providecommand{\bibinfo}[2]{#2}
\providecommand{\BIBentrySTDinterwordspacing}{\spaceskip=0pt\relax}
\providecommand{\BIBentryALTinterwordstretchfactor}{4}
\providecommand{\BIBentryALTinterwordspacing}{\spaceskip=\fontdimen2\font plus
\BIBentryALTinterwordstretchfactor\fontdimen3\font minus
  \fontdimen4\font\relax}
\providecommand{\BIBforeignlanguage}[2]{{%
\expandafter\ifx\csname l@#1\endcsname\relax
\typeout{** WARNING: IEEEtran.bst: No hyphenation pattern has been}%
\typeout{** loaded for the language `#1'. Using the pattern for}%
\typeout{** the default language instead.}%
\else
\language=\csname l@#1\endcsname
\fi
#2}}
\providecommand{\BIBdecl}{\relax}
\BIBdecl

\bibitem{early_work_1}
L.-H. Chen, Z.-H. Ling, L.-J. Liu, and L.-R. Dai, ``Voice conversion using deep
  neural networks with layer-wise generative training,'' \emph{IEEE/ACM
  Transactions on Audio, Speech, and Language Processing}, vol.~22, no.~12, pp.
  1859--1872, 2014.

\bibitem{early_work_2}
T.~Nakashika, T.~Takiguchi, and Y.~Ariki, ``Voice conversion using
  speaker-dependent conditional restricted boltzmann machine,'' \emph{EURASIP
  Journal on Audio, Speech, and Music Processing}, vol. 2015, no.~1, pp. 1--12,
  2015.

\bibitem{early_work_3}
S.~Desai, A.~Black, B.~Yegnanarayana, and K.~Prahallad, ``Spectral mapping
  using artificial neural networks for voice conversion,'' \emph{IEEE
  Transactions on Audio, Speech, and Language Processing}, vol.~18, pp. 954 --
  964, 08 2010.

\bibitem{early_work_7}
M.~Zhang, B.~Sisman, L.~Zhao, and H.~Li, ``Deepconversion: Voice conversion
  with limited parallel training data,'' in \emph{Speech Communication}, 2020,
  pp. 31--43.

\bibitem{early_work_8}
S.~H. {Mohammadi} and A.~{Kain}, ``Voice conversion using deep neural networks
  with speaker-independent pre-training,'' in \emph{SLT}.\hskip 1em plus 0.5em
  minus 0.4em\relax IEEE, 2014, pp. 19--23.

\bibitem{CycleGAN2017}
J.-Y. Zhu, T.~Park, P.~Isola, and A.~A. Efros, ``Unpaired image-to-image
  translation using cycle-consistent adversarial networks,'' in \emph{ICCV},
  2017, pp. 2223--2232.

\bibitem{CycleGAN-VC}
T.~Kaneko and H.~Kameoka, ``Cyclegan-vc: Non-parallel voice conversion using
  cycle-consistent adversarial networks,'' in \emph{EUSIPCO}.\hskip 1em plus
  0.5em minus 0.4em\relax IEEE, 2018, pp. 2100--2104.

\bibitem{CycleGAN-VC2}
T.~Kaneko, H.~Kameoka, K.~Tanaka, and N.~Hojo, ``Cyclegan-vc2: Improved
  cyclegan-based non-parallel voice conversion,'' in \emph{ICASSP}.\hskip 1em
  plus 0.5em minus 0.4em\relax IEEE, 2019, pp. 6820--6824.

\bibitem{kaneko2020CycleGAN-VC3}
T.~Kaneko, H.~Kameoka, K.~Tanaka, and N.~Hojo, ``Cyclegan-vc3: Examining and improving cyclegan-vcs for
  mel-spectrogram conversion,'' in \emph{Interspeech}, 2020, pp. 2017--2021.

\bibitem{StarGAN-VC2}
T.~Kaneko, H.~Kameoka, K.~Tanaka, and N.~Hojo, ``Stargan-vc2: Rethinking conditional methods for stargan-based voice
  conversion,'' \emph{Interspeech}, pp. 679--683, 2019.

\bibitem{StarGAN-VC}
H.~Kameoka, T.~Kaneko, K.~Tanaka, and N.~Hojo, ``Stargan-vc: Non-parallel
  many-to-many voice conversion using star generative adversarial networks,''
  in \emph{SLT}.\hskip 1em plus 0.5em minus 0.4em\relax IEEE, 2018, pp.
  266--273.

\bibitem{zhao2020unpaired}
Y.~Zhao, R.~Wu, and H.~Dong, ``Unpaired image-to-image translation using
  adversarial consistency loss,'' in \emph{ECCV}.\hskip 1em plus 0.5em minus
  0.4em\relax Springer, 2020, pp. 800--815.

\bibitem{autoVC}
K.~Qian, Y.~Zhang, S.~Chang, X.~Yang, and M.~Hasegawa-Johnson, ``{A}uto{VC}:
  Zero-shot voice style transfer with only autoencoder loss,'' in \emph{ICML},
  2019, pp. 5210--5219.

\bibitem{hsu2016voice}
C.-C. Hsu, H.-T. Hwang, Y.-C. Wu, Y.~Tsao, and H.-M. Wang, ``Voice conversion
  from non-parallel corpora using variational auto-encoder,'' in
  \emph{APSIPA}.\hskip 1em plus 0.5em minus 0.4em\relax IEEE, 2016, pp. 1--6.

\bibitem{VAE}
D.~P. Kingma and M.~Welling, ``Auto-encoding variational bayes,'' \emph{arXiv
  preprint arXiv:1312.6114}, 2013.

\bibitem{Representation_learning_CPC}
A.~v.~d. Oord, Y.~Li, and O.~Vinyals, ``Representation learning with
  contrastive predictive coding,'' \emph{arXiv preprint arXiv:1807.03748},
  2018.

\bibitem{tobing2019non}
P.~L. Tobing, Y.-C. Wu, T.~Hayashi, K.~Kobayashi, and T.~Toda, ``Non-parallel
  voice conversion with cyclic variational autoencoder,'' in
  \emph{Interspeech}, 2019, pp. 679--678.

\bibitem{VAE_CPC}
J.~Ebbers, M.~Kuhlmann, and R.~Haeb-Umbach, ``Adversarial contrastive
  predictive coding for unsupervised learning of disentangled
  representations,'' in \emph{ICASSP}.\hskip 1em plus 0.5em minus 0.4em\relax
  IEEE, 2021.

\bibitem{VQVAE_CPC}
B.~van Niekerk, L.~Nortje, and H.~Kamper, ``Vector-quantized neural networks
  for acoustic unit discovery in the zerospeech 2020 challenge,'' in
  \emph{Interspeech}, 2020, pp. 4836--4840.

\bibitem{VQ-VAE_paper}
A.~Van Den~Oord, O.~Vinyals \emph{et~al.}, ``Neural discrete representation
  learning,'' in \emph{NeurIPS}, 2017, pp. 6306--6315.

\bibitem{NCE_image}
T.~Park, A.~A. Efros, R.~Zhang, and J.-Y. Zhu, ``Contrastive learning for
  conditional image synthesis,'' in \emph{ECCV}, 2020.

\bibitem{adversarial_loss}
I.~Goodfellow, J.~Pouget-Abadie, M.~Mirza, B.~Xu, D.~Warde-Farley, S.~Ozair,
  A.~Courville, and Y.~Bengio, ``Generative adversarial nets,'' in
  \emph{NeurIPS}, 2014, pp. 2672--2680.

\bibitem{NCE}
M.~Gutmann and A.~Hyv{\"a}rinen, ``Noise-contrastive estimation: A new
  estimation principle for unnormalized statistical models,'' in
  \emph{AISTATS}, 2010, pp. 297--304.

\bibitem{SimCLR}
T.~Chen, S.~Kornblith, M.~Norouzi, and G.~E. Hinton, ``A simple framework for
  contrastive learning of visual representations,'' in \emph{ICML}, 2020.

\bibitem{veaux2016superseded}
C.~Veaux, J.~Yamagishi, and K.~MacDonald, ``Superseded-cstr vctk corpus:
  English multi-speaker corpus for cstr voice cloning toolkit,''
  \emph{University of Edinburgh. The Centre for Speech Technology Research
  (CSTR)}, 2016.

\bibitem{yamamoto2020parallel}
R.~Yamamoto, E.~Song, and J.-M. Kim, ``Parallel wavegan: A fast waveform
  generation model based on generative adversarial networks with
  multi-resolution spectrogram,'' in \emph{ICASSP}.\hskip 1em plus 0.5em minus
  0.4em\relax IEEE, 2020, pp. 6199--6203.

\bibitem{he2016deep}
K.~He, X.~Zhang, S.~Ren, and J.~Sun, ``Deep residual learning for image
  recognition,'' in \emph{CVPR}, 2016, pp. 770--778.

\bibitem{isola2017image}
P.~Isola, J.-Y. Zhu, T.~Zhou, and A.~A. Efros, ``Image-to-image translation
  with conditional adversarial networks,'' in \emph{CVPR}, 2017, pp.
  1125--1134.

\bibitem{loizou2011speech}
P.~C. Loizou, ``Speech quality assessment,'' in \emph{Multimedia Analysis,
  Processing and Communications}, 2011, pp. 623--654.

\bibitem{wan2018generalized}
L.~Wan, Q.~Wang, A.~Papir, and I.~L. Moreno, ``Generalized end-to-end loss for
  speaker verification,'' in \emph{ICASSP}.\hskip 1em plus 0.5em minus
  0.4em\relax IEEE, 2018, pp. 4879--4883.

\bibitem{jia2018transfer}
Y.~Jia, Y.~Zhang, R.~J. Weiss, Q.~Wang, J.~Shen, F.~Ren, Z.~Chen, P.~Nguyen,
  R.~Pang, I.~L. Moreno \emph{et~al.}, ``Transfer learning from speaker
  verification to multispeaker text-to-speech synthesis,'' in \emph{NeurIPS},
  2018, pp. 4480–--4490.

\bibitem{kumar2019melgan}
K.~Kumar, R.~Kumar, T.~de~Boissiere, L.~Gestin, W.~Z. Teoh, J.~Sotelo,
  A.~de~Br{\'e}bisson, Y.~Bengio, and A.~Courville, ``Melgan: Generative
  adversarial networks for conditional waveform synthesis,'' in \emph{NeurIPS},
  2019.

\end{thebibliography}

\end{document}